\documentclass{aa}
\usepackage{graphics}

\begin{document}

\title{A pseudo-planar, periodic-box formalism for modelling the outer evolution 
of structure in spherically expanding stellar winds
}

\titlerunning{Periodic box models of Wind Structure}

\authorrunning{Runacres and Owocki}

\author{M.C. Runacres \inst{1} \and S.P. Owocki \inst{2}}

\offprints{M.C. Runacres,
\email{Mark.Runacres@oma.be}}

\institute{Royal Observatory of Belgium, Ringlaan 3, B-1180 Brussel, Belgium
\and Bartol Research Institute, University of Delaware, Newark, DE 19716, USA}

\date{Received / Accepted}

\abstract{
We present an efficient technique to study the 1D evolution of
instability-generated structure in winds of hot stars out to very large
distances ($\sim 1000$ stellar radii). This technique makes use of our
previous finding
that external forces
play little r\^ole in the outer evolution of structure. Rather than 
evolving the entire wind, as is traditionally done, the technique focuses on a
representative portion of the structure and follows it as it moves out with
the flow. This requires the problem to be formulated in a moving reference
frame. The lack of Galilean invariance of the spherical equations of
hydrodynamics is circumvented by recasting them in a pseudo-planar form. By
applying the technique to a number of problems we show that it is fast and
accurate, and has considerable conceptual advantages. It is particularly
useful to test the dependence of solutions on the Galilean frame in which
they were obtained.
As an illustration, we show that, in a
one-dimensional approximation, the wind can remain structured out to 
distances of more than 1300 stellar radii from the central star.
\keywords{stars: early-type -- stars: mass-loss -- stars: winds, outflows
-- hydrodynamics -- instabilities}
}
\maketitle

\section{Introduction}
The line-driven stellar winds of hot, luminous, OB-type stars are subject to a 
strong line-deshadowing instability (e.g. Owocki \& Rybicki \cite{OR84}, 
Feldmeier \cite{Habitur}).
Hydrodynamical studies aimed at following the nonlinear evolution
of this deshadowing instability 
(Owocki, Castor \& Rybicki \cite{OCR},
Feldmeier et al \cite{FPP97},
Runacres \& Owocki \cite{Paper I} -- hereafter 
Paper~I) show that within a few stellar radii of the surface the flow 
becomes highly structured, with gas concentrated in dense clumps,
and pervaded by strong shocks.
To determine the full physical and observational significance of such 
structure, it is important to understand its subsequent development 
at scales beyond the few stellar radii covered in such
radiation-hydrodynamical simulations of its initial formation.
For example, for a star such as \object{$\zeta$~Pup}, nearly half of the 
observed thermal radio flux is understood to originate at distances of
more than $100\;R_{\ast}$. 
For non-thermal emitters such as
\object{Cyg~OB2~No.~9}, shocks are needed beyond $500\;R_{\ast}$ (Van Loo et al.
\cite{Van Loo etal}). For some stars (such as \object{$\zeta$~Pup}) it has
been suggested that a significant contribution to the
X-ray flux originates beyond $100\;R_{\ast}$ (Hillier et al \cite{Hillier etal}).
Recently, however, Kramer et al.
(\cite{Kramer etal}) have found fits to X-ray emission lines that indicate a
formation region much closer to the star ($r \la 5 \; R_{\ast}$).
Finally, 
Grosdidier et al. (\cite{Grosdidier+}) have suggested that some
of the structure found in ring-nebulae around Wolf-Rayet stars might be an
imprint of stellar wind structure.

These considerations demonstrate the importance of modelling the 
dynamical evolution of instability-generated structure out to very 
large distance scales of order $\sim 1000 \; R_{\ast}$.
Because of the computational expense of the
non-local integrals for calculating the radiative force central to the 
developing instability, full radiation hydrodynamical models have 
generally been limited to the distances of order $\sim 10 \; R_{\ast}$.
But in Paper~I we showed that such radiative forces become of 
negligible importance beyond distance of a few times $\sim 10 \;
R_{\ast}$, and so within a hybrid approach that switches to a much 
less costly, pure-hydrodynamical model of such outer regions, we were 
able to extend high-resolution, 1D instability models to distances 
of order $\sim 100 \; R_{\ast}$.
Unfortunately, even with this approach, extension to still larger 
scales again quickly becomes quite computationally expensive, since it
requires both a large number of depth points and a evolution of the 
model for the extended time required for advected structure to relax 
over such scales.

To address this problem, the present paper introduces a new
{\it pseudo-planar, moving periodic-box} approach.
This further reduces the complexity of the models by transforming the 
analysis into the mean rest frame of some representative section -- the box -- 
of the expanding flow structure.
Then, to account approximately for the secular radial changes associated 
with spherical expansion, the flow variables and their governing equations 
are recast in forms that resemble those for simple planar 
flow, the ``pseudo-planar'' form.
Finally, under the assumption that the chosen section represents a 
randomly characteristic sample of the flow structure, its internal 
evolution is then isolated by assuming periodic boundary conditions 
to link the inner and outer edges of the box.

In the following we first (\S \ref{sect:driven}) review the
radiatively driven models that serve as basic input to our approach, 
and then (\S \ref{sect:perbox}) formally introduce and develop the pseudo-planar, 
periodic box formalism.
We next (\S \ref{sect:test}) apply this to a number of test problems, 
and (\S \ref{sect:comparison}) compare the results with those of 
traditional, radiatively driven models.
In both sections, we devote particular attention to the effect of
advection and show that the ability to test the dependence of problems on the
Galilean frame is a major advantage of the periodic box model. 
After illustrating (\S \ref{sect:results}) the application of the method 
in a model with structure extending to $1300 \; R_{\ast}$, we conclude 
(\S \ref{sect:summary}) by summarizing the advantages, potential 
applications, and future extensions of a periodic-box approach.

\section{Radiatively driven models of outer wind
structure}\label{sect:driven}
In Paper~I we investigated 
structure up to a distance of $100 \;R_{\ast}$,
using Eulerian, one-dimensional, time-dependent hydrodynamical models that take
into account the instability of the driving. In these models, the material is
compressed into a sequence of narrow, dense shells, bounded by shocks.  These 
shells expand at roughly the sound speed as they move out at approximately 
the terminal velocity of the wind. There are supersonic velocity differences 
between individual shells, causing them to collide and form new shells.
The importance of similar shell-shell collisions for the production of X-rays
has 
already been pointed out by Feldmeier et al. (\cite{FPP97}). We found that these collisions 
effectively hinder the decay of the structure initiated in the inner
wind, so that the clumps can survive to substantial ($\ga 100 \;R_{\ast}$) distances.

In modelling the distant wind structure, it is necessary to maintain
a relatively fine grid spacing to resolve the often quite narrow dense clumps.
For a radiatively driven calculation, we use a grid of 10\,637 points. The
initial 1025 points have a spacing that increases linearly from 0.001 to 0.01
$R_{\ast}$ over the range $1-5\; R_{\ast}$. Beyond $5 \; R_{\ast}$ we use a constant grid spacing
of $0.01 \; R_{\ast}$. This is also the spacing of the box models.

Another important factor in the outer evolution of structure is the
energy balance in the wind.
The gas in our simulations cools both
adiabatically and radiatively. The combined effect of this cooling is balanced
by photoionisation heating from the star's ultraviolet radiation. We mimic the 
effect of radiative heating by setting a floor temperature, below which the 
temperature is not allowed to drop. The value of this floor temperature 
influences the expansion speed of the shells. For a low value of the floor
temperature, the dissipation of the structure will be slower. 
In Paper~I, the floor temperature was taken equal
to the effective temperature. This is a crude approximation and probably
results in a wind that is too warm. As a first step towards a more realistic
treatment of the energy balance, we used the detailed ionisation and thermal
equilibrium models by Drew (\cite{Drew}). These take into account the cooling by
heavy element lines and predict an outward decreasing temperature that quickly
reaches values substantially below the effective temperature. These
temperature profiles can be adequately approximated by the expression
\begin{equation}
\frac{T(r)}{T_{\rm eff}} =0.79 -0.51 \frac{v(r)}{v_{\infty}}
\end{equation}
(Bunn \& Drew \cite{Bunn&Drew}), which we used in the current models,
with the velocity assumed to follow the usual ``beta-law" form $v(r) = v_{\infty} 
(1-R_{\ast}/r)^{\beta}$ with exponent $\beta = 0.7$. 
Note that the temperature profile levels off at $0.28 \;T_{\rm eff}$.
Although this represents a modest improvement over simply taking the 
effective temperature, it is still quite unrealistic.
The Drew models were only calculated to $10 \;R_{\ast}$. Most
importantly, they assume a smooth outflow. There are numerous ways in which
the inhomogeneity of the outflow could influence the energy balance. 
A detailed investigation of the energy balance in a structured wind, however,
is beyond the present scope.

\begin{figure}
\resizebox{\hsize}{!}{\includegraphics{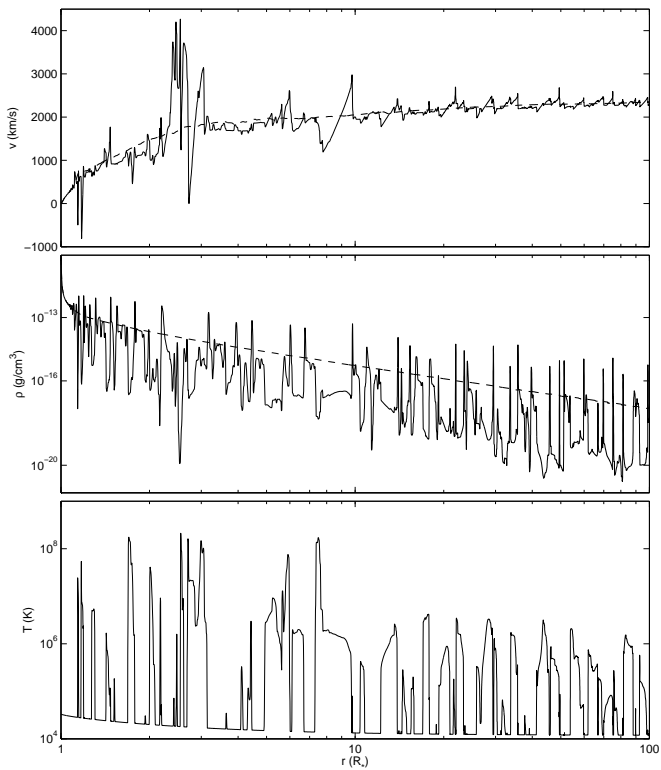}}
\caption{Velocity, density and temperature for a radiatively driven model.
The dashed lines in the two upper panels represent the time-averaged
variable.
  }
\label{fg:driven:snapshot}
\end{figure}

Finally, the amount of structure
initiated in the inner wind strongly depends on the line driving parameters, in
particular on the cut-off parameter $\kappa_{\rm max}$ that limits the maximum
line strength (Owocki, Castor \& Rybicki \cite{OCR}). For purely computational reasons,
this parameter is usually set to artificially low values. We found that, with the
relatively fine resolution of our calculations, it is possible to set this 
parameter to less artificial values. 
In the current paper, we have used 
$\kappa_{\rm max}=0.1\kappa_0$,
where the opacity constant $ \kappa_0$ is related to the actual 
strength of the strongest line. This is a factor of hundred larger than in
Paper~I. The effect is to include a number of strong
lines that become optically thin only for very large velocity gradients. This
allows for much stronger rarefactions and shocks. The resulting models are
extremely structured, as can be seen from the snapshot 
(Fig.~\ref{fg:driven:snapshot}).

In summary, we can say that the amount of
clumping in our simulations is largely determined by the grid spacing, the floor
temperature and $\kappa_{\rm max}$.
On the other hand, we find that the clumping does {\em not} depend on the 
radiative force beyond $\sim 30 \;R_{\ast}$ (Paper~I). This reduces the outer evolution of 
structure to a pure gasdynamical model. As the evaluation of the radiative
force dominates the computation time, this allows us to construct vastly
more economical models, which we present in the following section.

\section{A pseudo-planar moving periodic box formalism}\label{sect:perbox}
\subsection{Motivation and outline}
Even without the evaluation of the radiative force, the modelling of structure
out to {\em very} large distances ($\sim~1000 \;R_{\ast}$) is still expensive
and
the fine grid spacing needed to resolve the shells results in
impractically large files. While such models are marginally feasible in 1D, 
generalizing them to higher dimensions would be computationally prohibitive.
In this section, we present a technique that does not suffer from this
disadvantage, and has considerable conceptual benefits.

First, let it be noted that the structure generated by the instability, apart from being 
stochastic, is also quasi-regular in the sense that similar features are 
repeated over time. Therefore it is not necessary to keep track of the 
whole stellar wind during the duration of the simulation. It is enough 
to select a limited but representative portion of the structure, and follow this as 
it moves out with the flow. Following a portion of the wind entails
transforming the conservation equations to a moving reference frame. This is
not possible directly, as the spherical equations of hydrodynamics are not
invariant under a Galilean transformation. This problem can be circumvented by
rewriting the equations in a pseudo-planar form (see below).

A fundamental point in the present analysis is that
the time-dependent hydrodynamical variables can be viewed as having two
distinct components, varying on different scales. The first is a
rapidly oscillating variable, varying on
short spatial and temporal scales. The second is a slowly
varying function of radius, corresponding to the secular evolution
of the time-averaged variable.
As an example, consider the density in Fig.~\ref{fg:driven:snapshot}.
The gas in this simulation is
highly clumped, with density variations of several orders of magnitude
over less than a stellar radius. The {\em time-averaged} density however, is a
much more well-behaved function, decreasing steadily outward. Indeed, one of
the more reassuring properties of the models including the deshadowing
instability is that the mean
variables closely resemble the results from CAK theory. In our pseudo-planar
reformulation
of the hydrodynamical equations we try to absorb, as much as possible,
the secular evolution in the scaling of the variables.

\subsection{Pseudo-planar scaling}\label{pp scaling}

For the sake of clarity,
let us recapitulate the spherical equations of
hydrodynamics.  In Eulerian form, the one-dimensional (1D)
time-dependent equations for conservation of
mass, momentum, and energy are:
\begin{eqnarray} \label{eq:sp_euler}
\frac{\partial\rho}{\partial t} +%
\frac{1}{r^2}\frac{\partial(r^2 \rho v)}{\partial r}&=&0 \label{eq:sp_euler1}\\
\frac{\partial(\rho  v)}{\partial t}+%
\frac{1}{r^2}\frac{\partial(r^2 \rho v^2)}{\partial r}&=&%
-\frac{\partial p}{\partial r}
+f\label{eq:sp_euler2}\\
\frac{\partial e}{\partial t} +%
\frac{1}{r^2}\frac{\partial (r^2 e v)}{\partial r}&=&
- \frac{p}{r^2}\frac{\partial (r^2 v)}{\partial r}
- Q\label{eq:sp_euler3}.
\end{eqnarray}
Here 
$f$ is the sum of the
external forces (gravity and radiative driving) acting on the gas 
and $Q$ the power emitted by radiation per unit volume.
The internal energy density $e$ (in erg/cm$^3$) is related to the
pressure by the perfect gas law 
$p = (\gamma -1) e 
= \rho k T/(\mu m_{\rm H})$, 
which supplements the
set of equations.
The other symbols have their usual meaning. In this paper, we assume
a perfect monatomic gas and use $\gamma = 5/3$ , unless otherwise specified.

The first step in our reformulation of the equations of hydrodynamics 
is to scale out, as much as possible, the secular components of
the density, velocity and pressure.
From the continuity equation it can be seen that the secular
expansion of the gas at large distances from the star
(where we have reached the terminal velocity)
causes the density to fall of as $1/r^{2}$.
This suggests the following definition of the scaled density
$\tilde\rho$:
\begin{equation}
r_0^2 \tilde\rho = r^2 \rho,
\end{equation}
where $r_0$ is a fiducial radius.
As the mean temperature in the present models is almost constant, it is
logical to scale the pressure in the same manner as the density. The scaled 
internal energy then follows from the perfect gas law
$\tilde{p}=(\gamma-1)\tilde{e}$. The mean velocity in the outer wind is
roughly constant and need not be scaled.

Rewriting Eqs.~(\ref{eq:sp_euler1})-(\ref{eq:sp_euler3})
in terms of the scaled variables, we obtain
the following conservation equations:
\begin{eqnarray}\label{eq:pp_euler}
\frac{\partial\tilde\rho}{\partial t} +%
\frac{\partial \left(\tilde\rho v\right)}{\partial r} &=&0    \label{eq:scaled1}\\
\frac{\partial\tilde\rho v}{\partial t} +%
\frac{\partial \left(\tilde\rho  v^2 \right)}{\partial r} & =& %
-\frac{\partial \tilde p}%
{\partial r} + \tilde{f} +\frac{2\tilde p}{r}     \label{eq:scaled2}\\
\frac{\partial\tilde{e}}{\partial t} +%
\frac{\partial \left(\tilde{e} v\right)}{\partial r}  & =&  %
-\tilde{p}\frac{\partial v}{\partial r}
+\tilde{Q} -\frac{2\tilde{p}v}{r},                      \label{eq:scaled3}
\end{eqnarray}
where the external force and the heating rate have been scaled in the same 
way as the density and the pressure. These equations are {\em pseudo-planar}
in the sense that they describe the spherical problem but formally
resemble the {\em planar} equations of hydrodynamics. The only formal
difference with the planar equations are the geometric source terms appearing
in the momentum and energy equations. These express the secular evolution of
momentum and energy in spherical geometry. The outer wind is essentially
expanding at a constant velocity while the temperature is kept constant by the
balance between heating and cooling. The
geometric source term $2\tilde p/r$ in the momentum equation represents the
pressure gradient associated with this secular expansion. The geometric source
term $2\tilde p v/r$ in the energy equation expresses the secular adiabatic 
cooling of the gas. 

\subsection{Transformation to moving box}
Let us now transform the pseudo-planar equations from the stellar rest frame to
a frame that is moving at a constant velocity $v_0$. We refer to this moving
frame as the box.
We define the position coordinate $x$ within the box
by the Galilean transformation
\begin{equation}
 r=R_0 + v_0 t + x,
\end{equation}
where $R_0$ is the initial inner radius of the box.
For any variable
$\alpha$ depending on position and time
we can then convert from $\alpha(r,t)$ to $\alpha(x,t)$.
We denote the time-derivative at constant $x$ by ${\rm d/d}t$.
We then have
\begin{equation}
\frac{\rm d\alpha}{{\rm d} t} \equiv %
\left.\frac{\partial\alpha}{\partial t}\right|_x = %
\left.\frac{\partial\alpha}{\partial t}\right|_r + %
\left(\vec{v_0}\cdot\vec\nabla\right)\alpha
\end{equation}
If we further introduce the velocity $w$ with respect to the box velocity,
i.e. $v=v_0+w$, then we can rewrite the equations of hydrodynamics as
\begin{eqnarray}
\frac{\rm d\tilde\rho}{{\rm d} t}+
\frac{\partial \left(\tilde\rho w\right)}{\partial x} &=&0    \label{moving1}\\
\frac{\rm d\left(\tilde\rho w\right)}{{\rm d} t}+%
\frac{\partial \left(\tilde\rho  w^2\right)}{\partial x} & =& %
-\frac{\partial \tilde p}%
{\partial x} + \tilde{f}+\frac{2\tilde p}{R_0+v_0 t+x}   \label{moving2}\\
\frac{{\rm d}\tilde e}{{\rm d}t} +%
\frac{\partial\left(\tilde e w\right)}{\partial x}&=&%
-\tilde p \frac{\partial w}{\partial x}
+\tilde{Q}-\frac{2\tilde{p}(v_0+w)}{R_0+v_0 t+x}\label{moving3}
\end{eqnarray}

Finally, we impose periodic boundary conditions on the box. The box method
has a number of advantages that should be clear already. Not only is the 
number of grid points reduced with respect to a traditional
calculation where the whole wind is evolved, it is also not necessary to
increase the number of depth points if the structure needs to be followed
further. 
The computing time is less, due to the smaller number of
grid points. Furthermore, any explicit time-stepping method is subject to a
stability condition on the Courant number $C =  c \Delta t/\Delta x$, where
$c$ is the maximum propagation speed of information,
$\Delta t$ the time step and 
$\Delta x$ the grid spacing.
In order to be stable, 
$C$ must be smaller than one (this is the Courant-Friedrichs-Lewy condition),
and preferably a lot smaller.
Due to the slower speed at which features are
advected over the numerical grid, the condition on the Courant number
is less restrictive for the box model. The gain is substantial: 
permitted time steps increase from
$\sim \Delta x/v_\infty$ to a few times $\Delta x/a$, where 
$a$ is the sound speed. In the simulations presented here, 
$v_\infty/a \approx 150$.

\subsection{Effect of periodic boundary conditions}
\paragraph{General principles}
As a consequence of the periodic boundary conditions, structure is always
dragged along with the box. To ensure that the time scale over which the
structure evolves is comparable to the time-scale associated with its outward
movement, the speed of the box must be roughly the same as the terminal
velocity. This is clear from the fact that it is obviously not possible to
evolve structure cheaply by taking a very large box speed. Furthermore, as
shell collisions are very important for the evolution of structure (Paper~I),
the box should be big enough to contain more than one shell, even at late
times.
Finally, the periodic boundary conditions move features such as dense
shells from one side of the box to the other. The box method is 
inherently incapable of describing variations on scales beyond a box length. 
This means that the length of the box should be small compared to the distance
covered by the box.

\paragraph{Restriction of box size from energy equation}
A quantitative restriction on the box size can be derived from the secular
change of the energy density over the length of the box. Using the energy
equation (\ref{moving3}) we can write the relative change in the secular
component of the energy density over a box length $L$ as
\begin{equation}
\epsilon_{\rm e} \equiv \frac{1}{\tilde e}\int_0^{\tau} \frac{2 \tilde{p}(v_0+w) }{R_0 +v_0 t +x}dt,
\end{equation}
where $\tau = L/v_0$. If we assume that the change in $\tilde{p}$ is slow
compared to $1/t$ this gives
\begin{equation}
\epsilon_{\rm e} = 2 (\gamma-1) \ln\left(1+\frac{L}{R_0+x}\right).
\end{equation}
Using $x=L/2$ and $R_0 = 50 R_*$
the requirement $\epsilon_{\rm e} <0.25$ gives $L < 11.5 R_*$.

\paragraph{Restriction from clumping factor}
A different measure of the effect of periodic boundary conditions is given by
their influence on the clumping factor. When a shell crosses the boundary of
the box, it effectively jumps a distance $L$ in the stellar rest frame. If
the clumping factor changes over length scales $\sim L$, this 
introduces an error, which we can estimate in the following simplified model.
Let us assume a single shell in the box, with
an expansion speed $a_{\rm e}$. If the shell has a width
$l_{\rm L}$ at the left side of the box then it has a width $l_{\rm L} +
\Delta l$ at the right side of the box, where $\Delta l = a_{\rm e}
L/v_{\infty}$. Note that the relevant time-scale for the expansion is {\em
not} the time needed for the shell to cross the box, but the much shorter
time $L/v_0$ needed for the shell to move a distance $L$ in the stellar rest
frame. We can then write the relative change in the clumping factor caused by
applying the periodic boundary conditions as
\begin{equation}\label{eq:clumping error}
\epsilon_{\rm cl} \equiv \frac{f_{\rm cl,L} -f_{\rm cl, R}}{f_{\rm cl, R}}
= \frac{a_{\rm e}}{v_\infty}\frac{L}{l_{\rm L}},
\end{equation}
where we have used the fact that the clumping factor can be approximated by 
the inverse of the volume filling factor, 
which in turn is the fractional width occupied by the shell (see e.g.
Paper~I). This equation
shows that the error caused by applying periodic boundary conditions is large
if the shells expand fast, the outflow velocity is small, the box is big or
the shells are narrow. This last possibility is explained by the fact 
that for a given $L, a_{\rm e}$ and $v_{\infty}$, 
the expansion $\Delta l$ is constant and
therefore relatively more important for narrow shells.
From the models
presented later in this paper (Sect.~\ref{sect:results}),
we can derive typical values to use
in Eq.~(\ref{eq:clumping error}). At $\approx 170 R_*$, where the first shell crosses
the box (Fig.~\ref{fg:box:image}),
the clumping factor is less than 10. For an expansion speed of 50
km/s and a terminal speed of 2300 km/s we have $\epsilon_{\rm cl} \approx
20\%$, i.e.
for the model parameters used in this paper,
the error on the clumping factor caused by applying the periodic boundary
condition is less than $20\%$. The assumption of a single shell is not a
limitation, because the above analysis can be repeated for any number of
shells to obtain the same result.

\subsection{Implementation}
We have implemented the pseudo-planar periodic box technique in VH-1, a
hydrodynamics programme (or {\em hydrocode}) developed at the University of
Virginia (Blondin, personal communication). 
This code solves the  Eulerian equations of hydrodynamics by a
Lagrangian remap (LR) method, which first solves the Lagrangian equations of
hydrodynamics and then maps the updated quantities back onto the Eulerian
grid. 
A major challenge in Eulerian hydrodynamics, even in their LR incarnation,
is to obtain reliable estimates of time-averaged quantities such as pressure
and density at zone interfaces. All quantities in VH-1 are zone-centred, i.e.
VH-1 does not use a staggered mesh. 
To determine the time averages,
a Riemann problem (Zel'dovich \& Raizer \cite{Zeldovich&Raizer}) is set up, and solved,
at each zone interface. A parabolic interpolation is used
to provide an accurate guess to the values on either side of the interface. 
This combination of the parabolic interpolation and the Riemann solver is
referred to as the piecewise parabolic method 
(PPM; Colella \& Woodward \cite{CW84}) and is the heart of VH-1. 
The details of the implementation are described in Paper I, the obvious
exception being that the radiative force need not be evaluated in 
the present work. 

\section{Test problems}\label{sect:test}
\subsection{Uniformly expanding sphere}
A straightforward problem to test the pseudo-planar approach is that of an
expanding sphere with spatially constant density and pressure, and a radially
increasing velocity (see e.g. Blondin \& Lufkin \cite{BL93}). For 
an initial condition with density $\rho_0$, pressure
$p_0$ and velocity $v=r/t_0$ this problem has the analytic solution 
\begin{eqnarray}
\rho = \rho_0\left(\frac{t_0}{t}\right)^3,\qquad
v = \frac{r}{t},\qquad
p    = p_0   \left(\frac{t_0}{t}\right)^{3\gamma}.
\end{eqnarray}
This test problem was used by Blondin \& Lufkin to illustrate the use of
geometry corrections to
minimise advection errors near the origin of a curvilinear coordinate system.
For the present
simulation, we have used the version of VH-1 available from the North
Carolina State University website. This version does not include the
geometry corrections. To avoid problems with the derivation of the
internal energy from remapped quantities, Blondin \& Lufkin removed the
pressure gradient from the momentum equation. For the pseudo-planar method,
this would also remove the geometrical source term from Eq.~(\ref{eq:scaled2}), thus 
greatly reducing the stringency of our test. We therefore include the
pressure gradient in our calculations.

\begin{figure}
\resizebox{\hsize}{!}{\includegraphics{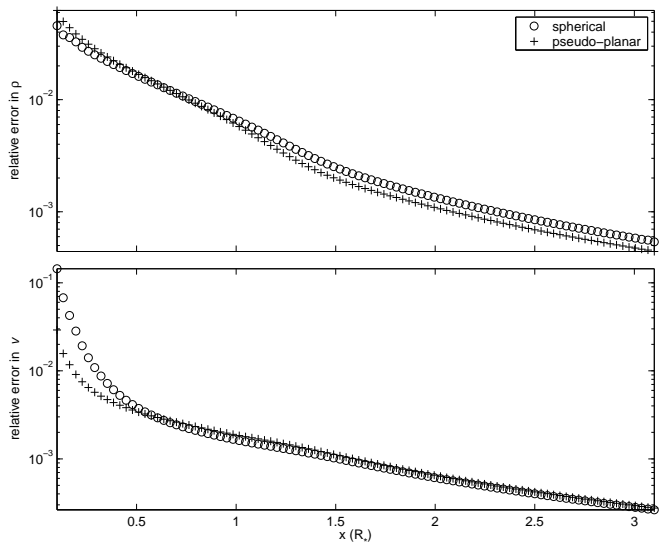}}
\caption{Relative difference between the analytic solution and the computer
simulation for the uniformly expanding sphere test problem. Upper panel:
density. Lower panel: velocity.
  }
\label{fg:Lufkin&Hawley}
\end{figure}

As a point of reference, we choose an initial condition that produces
physical parameters typical for the outer wind of a hot star: 
\begin{equation}
t_0 = 10^5 \mathrm{s},\quad
\rho_0 = 10^{-16}\mathrm{g/cm^3},\quad
p_0 = 5 \times 10^{-4}\mathrm{dyn/cm^2}.
\end{equation}
The spatial grid has 96 points and extends from 0.1 to 3.1 $R_{\ast}$, where 
$R_{\ast} = 19 \; R_{\sun}$. We compare both a spherical model, calculated by solving
Eqs.~(\ref{eq:sp_euler1})-(\ref{eq:sp_euler3}),
and a pseudo-planar model in a stationary box, calculated by solving 
Eqs.~(\ref{eq:scaled1})-(\ref{eq:scaled3}),
with the exact analytical solution. 
The Courant number is 0.25.
The boundary conditions were chosen so as to impose a
zero gradient on density and pressure, and a constant gradient on the
velocity.  In the case of the pseudo-planar method, 
the scaled variables (e.g. $\tilde\rho$) are implicitly scaled back to their physical
counterpart ($\rho$) before imposing the boundary condition. This is
important when comparing the two methods, as a difference in the quality of
the boundary conditions would swamp the relatively small errors intrinsic to the
method.
Figure~\ref{fg:Lufkin&Hawley} shows that the pseudo-planar model 
performs marginally better than the spherical model. 
This is gratifying, as the test is unfavourable to the pseudo-planar method, which has to
produce a flat physical density and pressure by evolving a curved rescaled
density and pressure. 

\subsection{A moving shock tube in planar geometry}
The Sod shock tube (Sod \cite{Sod}) has become a standard test for all
hydrodynamics codes. It is a special case of the Riemann problem, describing
the evolution of an arbitrary discontinuity (Zel'dovich \& Raizer 
\cite{Zeldovich&Raizer}). A hypothetical membrane separates two regions of
uniform density, pressure and velocity, where at least one of these
quantities is different on either side of the membrane. At $t=0$, the membrane
is removed and the jump discontinuity breaks up into a shock wave, a contact
discontinuity and a rarefaction wave. In planar geometry, this problem has a
semi-analytical solution (Sod \cite{Sod}, Laney \cite{Laney}). The standard 
problem presented by Sod,
with the gas at rest at both sides of the membrane, a density ratio of 8 and a
pressure ratio of 10, is quite unchallenging for any modern hydrocode. The
physical problems to which these codes are applied generally present more
extreme conditions. 

In radiatively driven stellar winds, density contrasts are often orders of
magnitude larger than in the standard Sod problem, while the gas moves over
the grid at supersonic speeds. We therefore propose a shock tube with a
density ratio of 800 and a pressure ratio of 1000, with all of the gas
initially moving at the same supersonic speed. By Galilean invariance, the
solution should not depend on this initial speed (except for an obvious
displacement).
On one side of the membrane, we have $\rho_l=100$,
$p_l=100$, $u_l=60$, on the other $\rho_r=0.125$, $p_r=0.1$,
$u_r=60$. Furthermore, to highlight possible differences between
forward and reverse waves, we set two such shock tubes back to back,
so that we have shocks, contact discontinuities and rarefaction waves
running in both directions. The 
initial situation is a layer of dense, high-pressure gas separated by two
membranes from the surrounding sparse, low-pressure gas, with everything
moving at the same speed. The original thickness of the dense layer is 0.2.

\begin{figure}
\resizebox{\hsize}{!}{\includegraphics{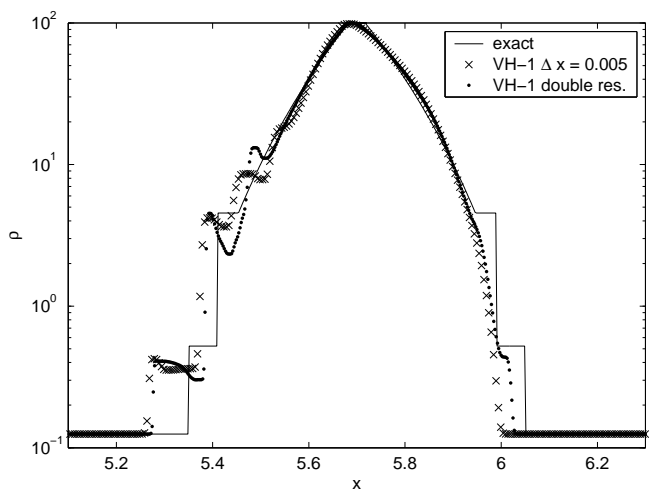}}
\caption{
A VH-1 solution of the density for the moving shock tube problem. 
The exact solution (solid line) is compared to a VH-1 simulation with
resolution $\Delta x = 0.005$ (crosses) and $\Delta x = 0.0025$ (dots).
}
\label{fg:mSod:VH1}
\end{figure}

The initial speed is more than fifty times the
adiabatic sound speed (the value of the adiabatic constant in this problem is
1.4). This is a more challenging test than the standard Sod
problem, but by no means more extreme than the conditions in a stellar wind.
In Fig.~\ref{fg:mSod:VH1} we compare the 
analytical solution (solid line) with the numerical solution using VH-1
in planar geometry,
with a step size of 0.005 (crosses) and a Courant number of 0.25.
The results are quite dismal, particularly for the inward
rarefaction, which shows a lot of unphysical substructure. Note also that the
inward running shock is not at the correct position. 

This substructure is present for a wide range of code parameters. 
Increasing the resolution by a factor of two
(dots) gives an unsubstantial improvement, as do variations in the
restriction on the Courant number (we repeated the calculation with 
Courant numbers between 0.1 and 0.6 and
found similar results). 
The artefacts occur regardless of whether the
internal energy is remapped separately or derived from the remapped total
energy (see below). They also occur for a wide range of shock flattening
parameters (see below)

In hot-star wind simulations, 
the internal energy is an ill-conditioned part of the total energy (Feldmeier
\cite{Feldmeier95}). We found that deriving the internal energy
from the remapped total
energy can result in artificial spikes of low temperature. To alleviate this
problem one can remap both the internal and total energy, and derive the
internal energy from the total
energy only when the Mach number of the flow is sufficiently low (Blondin,
personal communication). We found that, though this fix is important in our
radiatively driven calculations, it does not help with the artefacts shown in
Fig.~\ref{fg:mSod:VH1} . Indeed, the models shown includes this fix, which at $M=60$ means
that the internal energy is never determined using the total energy. When the
total energy is used, the results are worse.

We find that shock flattening (Colella
\& Woodward \cite{CW84}) does little to improve the results. This technique
flattens the interpolating parabola in the vicinity of a shock. In the
calculations presented here we have used the flattening parameters
 $\omega^{(1)}  = 0.75$,
 $\omega^{(2)}  = 5.0$ and
 $\epsilon = 0.33$.  
Even with extreme shock flattening
($\omega^{(1)}  = 0$,
 $\omega^{(2)}  = 10.0$ and
 $\epsilon = 0$) the artefacts do not disappear, and the amount of smearing
at the shocks is unacceptable. Flattening is usually only applied in the
Lagrangian hydro step. Extending flattening to the remap step doesn't improve
the results substantially.
If the interpolating function is totally flattened in very grid point (i.e.
by setting the flattening coefficient $f_i =1$ for all $i$), VH-1
reproduces the Godunov scheme. In this case, all features in the test problem
are completely wiped out, because the scheme is too dissipative. By trial
and error we find that  setting the flattening coefficient $f_i =0.025$ for
all $i$ gives acceptable results.

\begin{figure}
\resizebox{\hsize}{!}{\includegraphics{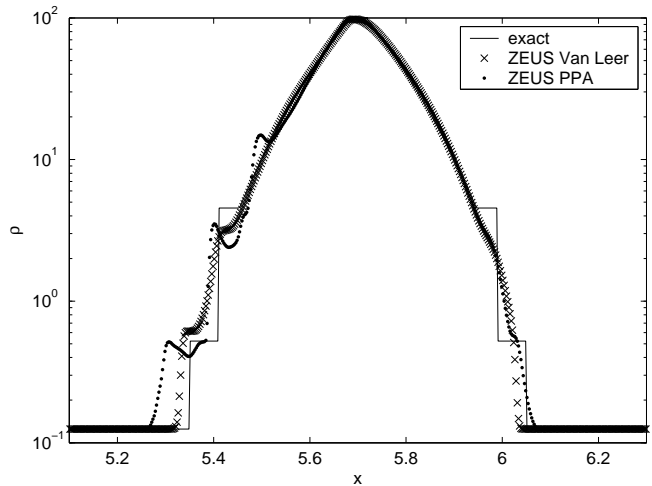}}
\caption{
Two ZEUS solutions of the moving shock tube problem. 
The exact solution (solid line) is compared to a ZEUS calculation using the
Van Leer scheme (crosses) and a calculation using PPA (dots).
}
\label{fg:mSod:Zeus}
\end{figure}

As a comparison, we ran the same simulation
using ZEUS, a widely used MHD code developed by Stone \& Norman
(\cite{SN92}). The numerical algorithms used to solve the Eulerian equations 
in ZEUS are quite different from those adopted in VH-1. It does not use a
Lagrangian remap, but directly solves the Eulerian equations. The solution 
is split into two parts: a {\em source} step and an {\em advection} step.
Although a staggered mesh is used, interpolations are still needed to
determine the time-averaged values of variables at zone interfaces. ZEUS
provides a second-order method (Van Leer) and a third-order method (PPA, for
piecewise parabolic advection). Note that PPA is only half of PPM, as it
incorporates the parabolic interpolation, but not the Riemann solver.
We have applied both the Van Leer scheme and PPA to the moving shock tube
problem. Figure~\ref{fg:mSod:Zeus} shows that the Van Leer scheme (crosses)
does much
\begin{figure}
\resizebox{\hsize}{!}{\includegraphics{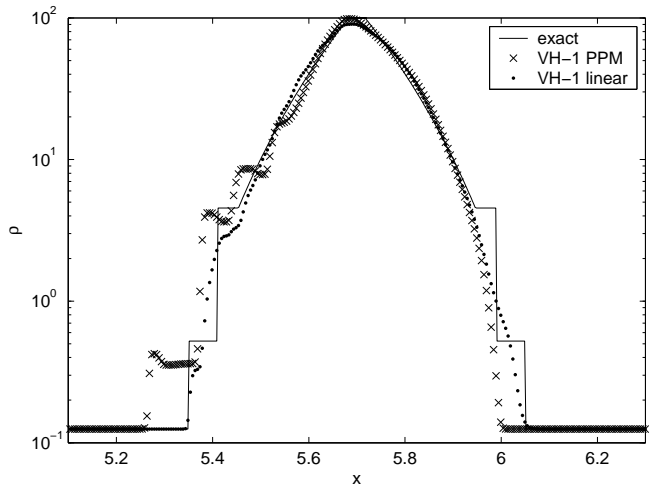}}
\caption{
Two VH-1 solutions of the moving shock tube problem. 
The exact solution (solid line) is compared to a a standard VH-1 calculation using
PPM (crosses) and a calculation using linear interpolation (dots).
}
\label{fg:mSod:lin}
\end{figure}
better on this problem than VH-1, while PPA (dots) produces the same kind
of unphysical substructure. 
The resolution used in this calculation is $\Delta x = 0.0025$.
At a resolution of 0.005 (not shown on figure),
the Van Leer scheme doesn't manage to capture the shock and the contact discontinuity,
but doesn't produce any spurious features either. 
Given the differences between the two codes,
it is quite surprising that the artefacts in VH-1 and ZEUS PPA are so similar.
It strongly suggests that they are the result of the parabolic interpolation
scheme used in both codes. To confirm this, we artificially lowered the order
of VH-1 by setting the quadratic term in the interpolating function to zero.
The results for the linear interpolation scheme are much better than for the
higher order PPM (Fig.~\ref{fg:mSod:lin}). 
The difference is due to errors in the remap step. This can
be seen from the fact that using linear  interpolation in the remap step only
gives essentially the same results.

The artificial substructure in this test problem
is similar to the rarefaction shocks found
by Falle (\cite{Falle}) in a ZEUS test calculation of a different Riemann
problem, using the Van Leer scheme. We have found, as expected, that VH-1
performs very well on Falle's test problem and that using PPA in ZEUS also
produces artefacts. 

\begin{figure}
\resizebox{\hsize}{!}{\includegraphics{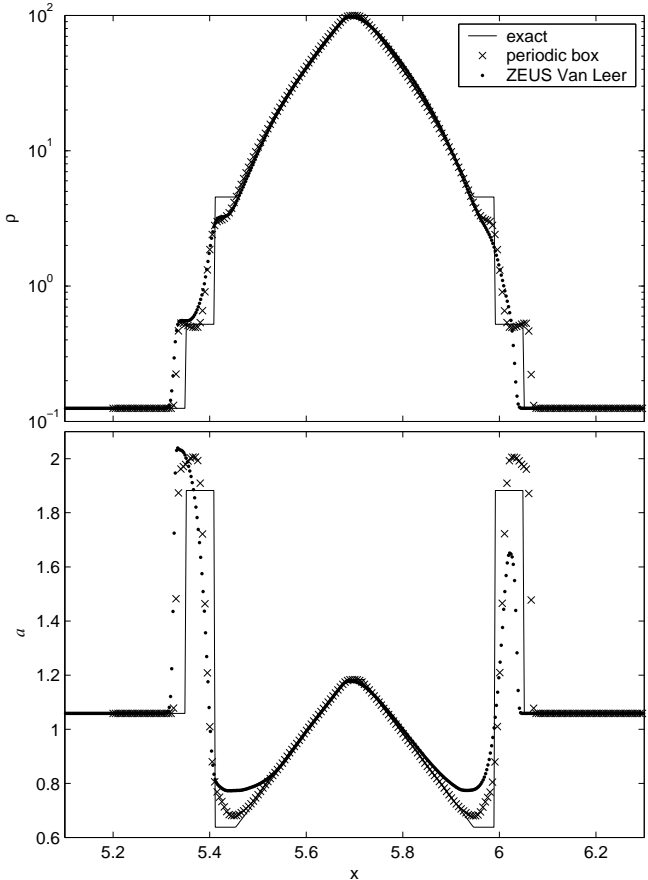}}
\caption{Density and sound speed for a periodic box solution of the moving
shock tube problem. The periodic box method (crosses) is compared to the
exact solution (solid line) and the ZEUS Van Leer method (dots).
}
\label{fg:mSod:pseudo-planar}
\end{figure}

Figure~\ref{fg:mSod:pseudo-planar} shows the density and the isothermal
sound speed for a VH-1 simulation using the pseudo-planar method. 
The pseudo-planar method, by its very nature, can only solve problems in a 
{\em spherical} geometry.
Planar geometry was mimicked by taking a very large initial position
of the box.
The box velocity was taken equal to the initial speed of the gas. 
The pseudo-planar method (crosses) is compared to the exact solution
and to the ZEUS Van Leer simulation described above. 
It is
clear that the pseudo-planar method performs very well on this test problem.
This shows that the main stumbling block for VH-1 and ZEUS PPA
is the supersonic velocity
with which the features move over the grid. Indeed, any box velocity reducing
this velocity to less extremely supersonic values gives adequate results.
With minor modifications, VH-1 can also be used as a pure Lagrangian code. In
this mode, the results are of the same quality as those of the periodic box
technique, which corroborates the above conclusion. 

In summary, we can say that the combination of parabolic interpolation and
highly supersonic advection leads to unphysical substructure and incorrect
shock speeds. The results can be improved by removing the advection from the
scheme (e.g. by using the periodic box technique) or by avoiding parabolic
interpolation. The lesson to be drawn from these experiments is that for
problems that are rapidly advected over the computational grid, higher order
schemes such as PPM do not necessarily give more accurate results than
lower-order schemes.

\section{Application to stellar winds}\label{sect:comparison}
\subsection{Initial condition}
To apply the moving periodic box simulation we select as an initial condition
a region from the snapshot shown in Fig.~\ref{fg:driven:snapshot}. The region
is chosen so that it contains a sufficiently large number of shells. This is
necessary to be able to evolve a model over a long period of time,
as the evolution of the structure is largely determined by the
fact that shells in the box collide and form denser shells
as the box
moves outward, counteracting their pressure-expansion. If only a single shell 
remains, the calculation becomes meaningless as there is obviously no 
opportunity for further collisions.
The boundaries of the box were set to generate a periodic velocity. The
density and pressure were then slightly modified near the inner boundary
of the box, to ensure periodicity in the corresponding {\em scaled} variables
and thereby avoid introducing any additional discontinuities. In practice,
the initial condition is the region $r= 46.1~-~94 \; R_{\ast}$ on the snapshot. 
The scaled density
is made periodic by introducing a linear ``correction" 
between the inner boundary and some $r_1$, so that the modified scaled
density is the same at the left and right boundaries and the correction
vanishes at $r_1$. The pressure is made periodic in the same manner.
Points beyond $r_1$ are not modified. In the simulations shown 
$r_1 = 49.2 \; R_{\ast}$.

\subsection{Comparison with radiatively driven model}
\subsubsection{Stationary periodic box}
\begin{figure}
\resizebox{\hsize}{!}{\includegraphics{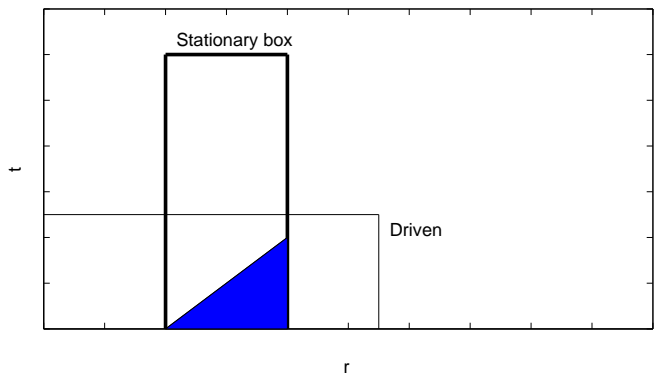}}
\caption{ Regions in the space-time plane covered by a full radiatively driven
model and a {\em stationary} periodic box model. The dark triangle indicates
the area in space-time covered by both models, i.e. the area where results of
the models can be compared
}
\label{fg:box:stationary}
\end{figure}

To better evaluate possible differences between a moving periodic box model
and the standard radiatively driven model, we first apply the pseudo-planar
equations in a {\em stationary} periodic box. We can compare the stationary
periodic box model with the driven model only in  a limited region of the $r,t$
plane. This is illustrated in Fig.~\ref{fg:box:stationary}. The wide
box represents the domain in space and time covered by a radiatively driven
model. The tall box represents the domain covered by a stationary box
model. In principle, the two models can be compared in the square region
where they overlap. However, the two models use different boundary
conditions, so the comparison is only meaningful in the region that
depends exclusively on the initial condition and not on the boundary
conditions. This region is the filled triangle in 
Fig.~\ref{fg:box:stationary}. The upper edge of this triangle is in fact
a $C_+$ characteristic (see e.g. Zel'dovich \& Raizer \cite{Zeldovich&Raizer}),
along which 
a signal from the inner boundary condition is carried outward. 
It need not be a straight line.
In the context of the present paper it can be viewed as the outer edge of an
expanding shell and the upshot of the above is that we can
only compare the box model with the driven model for shells that have not
crossed the boundary of the box.
 
\begin{figure}
\resizebox{\hsize}{!}{\includegraphics{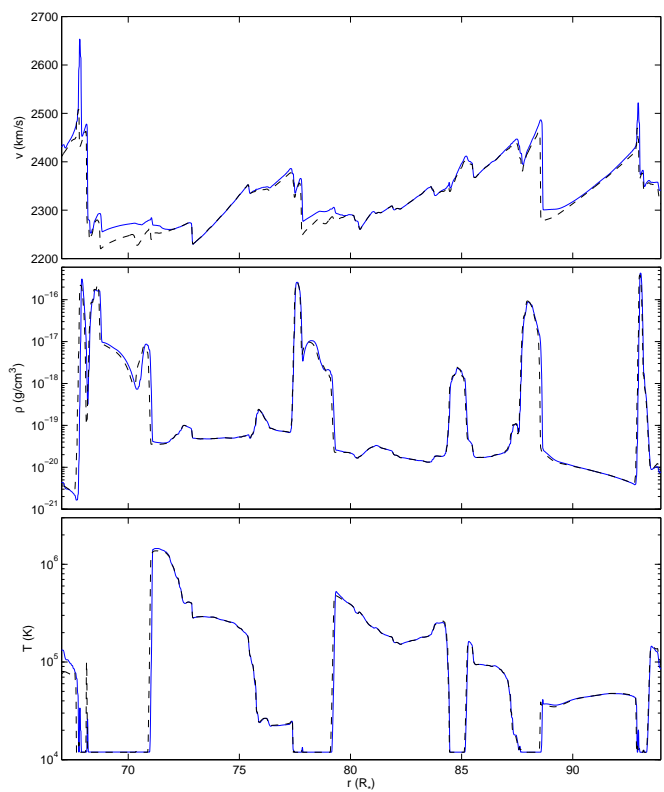}}
\caption{
Snapshot of velocity, density and temperature at 100 ksec for a
radiatively driven model (solid line) and a {\em stationary} periodic box
model (dashed line)
  }
\label{fg:box:compare:stationary}
\end{figure}

Figure~\ref{fg:box:compare:stationary} shows a snapshot at 100 ksec, for a  
stationary periodic box model (solid line) and a radiatively driven model
(dashed line). We show only
the region in $r$ where the comparison is meaningful. It is clear that the
two models agree very closely. The small differences are
most obvious in the velocity plot (upper panel) and appear to be due to the 
residual level of radiative driving beyond $45 \; R_{\ast}$.

\subsubsection{Moving periodic box}
\begin{figure}
\resizebox{\hsize}{!}{\includegraphics{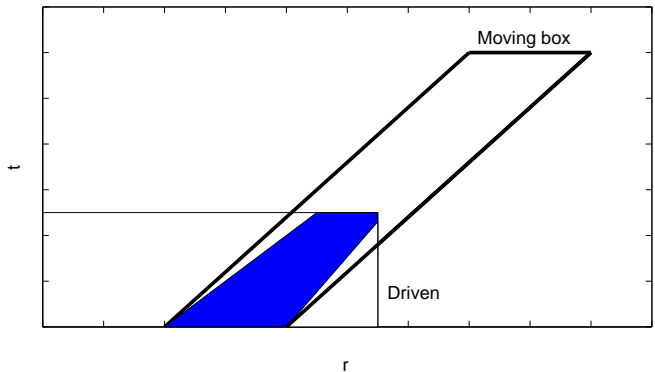}}
\caption{
Regions in the space-time plane covered by a full radiatively driven
model and a {\em moving} periodic box model. The dark trapezium indicates
the area in space-time covered by both models, i.e. the area where results of
the models can be compared
  }
\label{fg:box:moving}
\end{figure}

Let us now compare the results for a {\em moving} periodic box model with a
radiatively driven model. Figure~\ref{fg:box:moving} shows the domains in the
r,t plane covered by the moving box model and the radiatively driven model.
The region (filled) where the two can be meaningfully compared is 
somewhat larger 
than for the stationary box, but still limited, this time by both $C_+$ and
$C_-$ characteristics. Again, the meaningful region can be viewed as those
shells that have not crossed the boundary of the box.

\begin{figure}
\resizebox{\hsize}{!}{\includegraphics{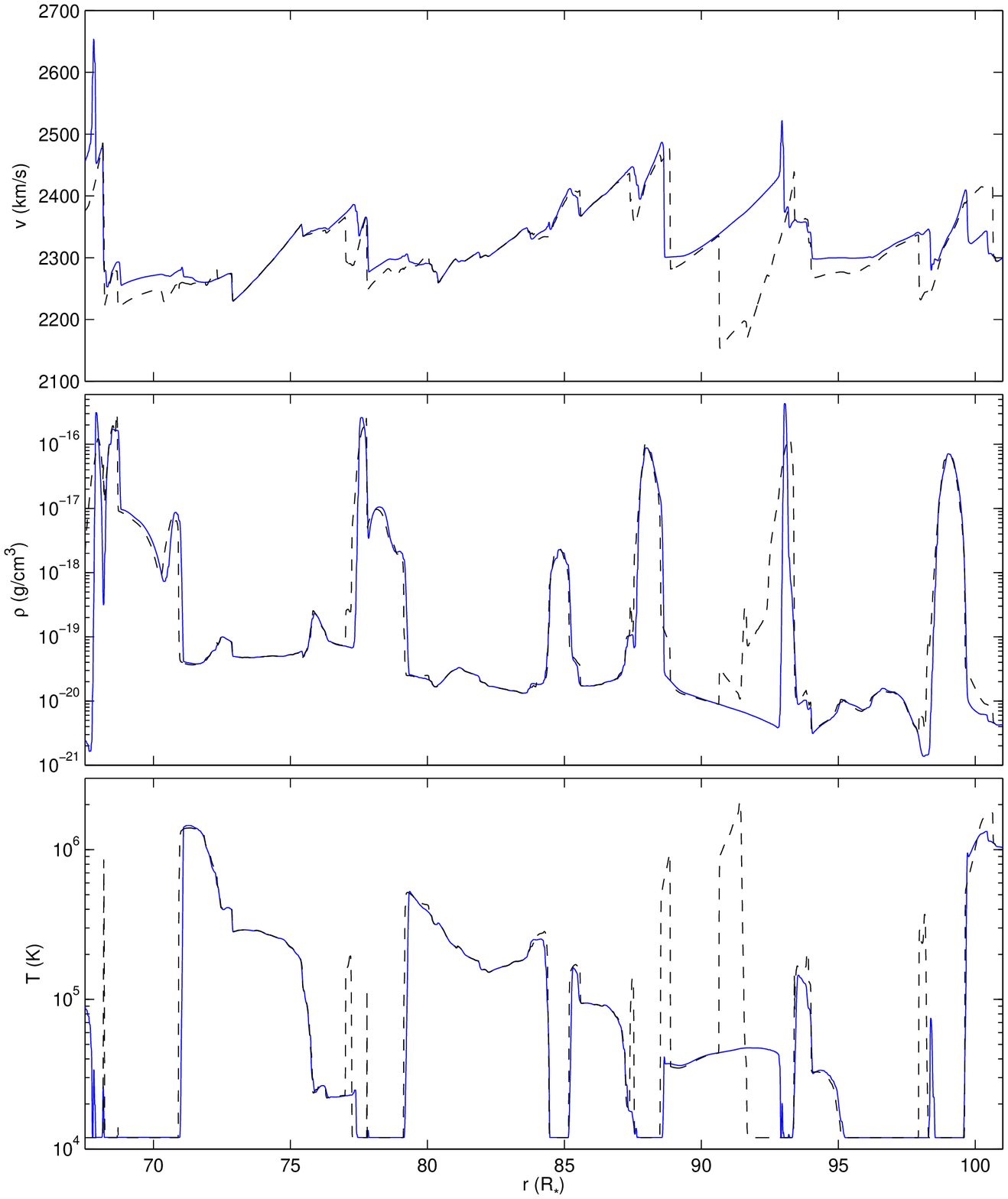}}
\caption{
Snapshot of velocity, density and temperature at 100 ksec for a
radiatively driven model (solid line) and a {\em moving} periodic box
model (dashed line)
  }
\label{fg:box:compare:moving}
\end{figure}

Figure~\ref{fg:box:compare:moving} shows a snapshot at 100 ksec, for a  
moving periodic box model (solid line) and a radiatively driven model
(dashed line). It is clear that the agreement is less than perfect.
In particular, the temperature of the moving box model shows features that are
not present in the driven calculation. Both models have broad regions of
rarefied hot gas (such as the one extending from 79 to 84.5 $R_{\ast}$). This is
gas that has been previously heated and has remained hot due to the
inefficiency of radiative cooling at low densities. (We recall that we solve
the energy equation including radiative cooling, using the Raymond et al.
[\cite{RCS}] cooling curve). In addition to these
broad warm regions, the periodic box model has narrow regions of gas heated
by nearby shocks. The most conspicuous pair of such narrow, hot regions is
centred around 90 $R_{\ast}$. These narrow regions are not present in
the case of a {\em stationary} periodic box. They appear due to the Galilean
transformation which is the only difference between the moving and stationary
periodic box calculations.  These narrow regions are perfectly physical:
the left region is heated by the forward shock at
$\approx 89 \; R_{\ast}$, the right region by the reverse shock at $90.5 \; R_{\ast}$. 
Their
narrowness is explained by the fact that the shocks have only been able to 
heat the gas during a 100 ksec time interval, and is consistent with the
velocity at which the gas flows out of the shock.  It is
in fact their absence in the radiatively driven calculation that is
disturbing. As gas passes through a shock, part of its kinetic energy is
converted into internal energy.  As it moves away from the shock,
the shock-heated gas cools by emitting radiation. A useful expression for the
length of the cooling zone was given by Feldmeier (\cite{Feldmeier95}). Using
his Eq. (A9) we find that the cooling length for even the weakest outer-wind
shocks is well-resolved by the numerical grid. The cooling length for the
relatively strong ($\chi =3.7$) reverse shock at $90.5 \; R_{\ast}$ in 
Fig.~\ref{fg:box:compare:moving} is a huge $400\; R_{\ast}$!

The absence of hot gas behind the shocks of the radiatively driven and
stationary periodic box model is a manifestation of the
``collapse" of cooling zones. This is a problem that has plagued all
hydrodynamical simulations of hot-star winds that include energy balance. It
has been attributed to a global thermal instability leading to an oscillation
of the width of the cooling zone (Feldmeier \cite{Feldmeier95}) and to
advective diffusion (Cooper \cite{CooperPhD}). Advective diffusion is caused
by the combination of advection with the typical shape of the cooling curve.
Between $10^5$ and $10^7$~K, this curve falls off as $T^{-1/2}$. When a steep
temperature feature is advected over the grid, it is inevitably smeared by
numerical diffusion. Due to the $T^{-1/2}$-dependence of the cooling, the
cold side of the smeared-out feature cools more rapidly than the warm side.
This effectively steepens the feature again, but also makes it narrower. In this
way, the numerical scheme takes a bite out of the cooling zone at every
time-step, until it is eaten away completely. Our results suggest that, at least
in the outer wind, advective diffusion is the cause of the collapse of
the cooling zones.

\section{Results for an example application}\label{sect:results}

\begin{figure}
\resizebox{\hsize}{!}{\includegraphics{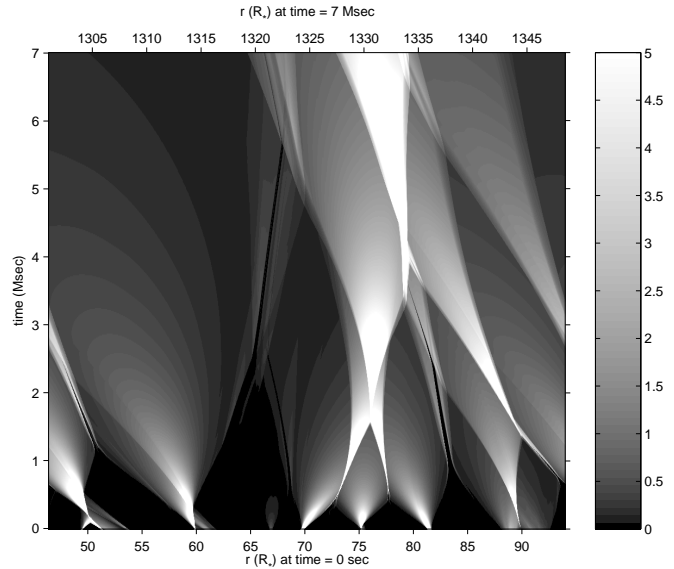}}
\caption{Greyscale image of the density contrast (density divided by mean
density) as a function 
of space and time. The intensity scale has been truncated to highlight the
kinematics of the shells.
  }
\label{fg:box:image}
\end{figure}
Let us now apply the periodic box technique to the problem it was 
designed for: the
evolution of stellar wind structure out to very large distances. In
Fig.~\ref{fg:box:image} we show the evolution of the density contrast
(density divided by mean density) as a greyscale image. During the
7 Msec interval shown, the box moves from 50 to 1300 $R_{\ast}$.
This figure shows the considerable conceptual advantage of the periodic box
method. By following a fraction of the wind as it moves away from the star,
the relevant physical mechanisms (pressure expansion and shell collisions)
are immediately apparent. Similar greyscale images can be made for the other
variables. 

\begin{figure}
\resizebox{\hsize}{!}{\includegraphics{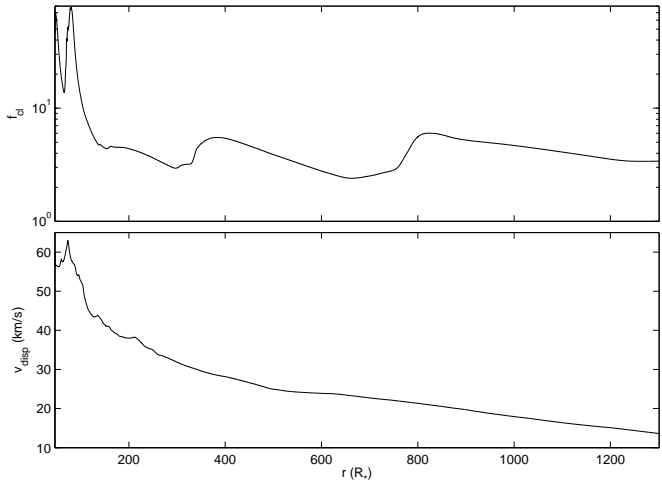}}
\caption{Clumping factor and velocity dispersion as a function of radius for
the box model.
  }
\label{fg:box:stat}
\end{figure}

\begin{figure}
\resizebox{\hsize}{!}{\includegraphics{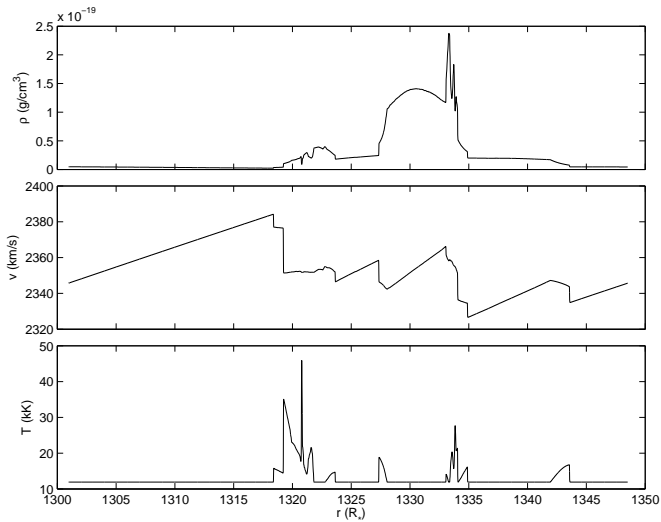}}
\caption{Snapshot of density, velocity and temperature at the end of the box
simulation (7 Msec).
  }
\label{fg:box:final}
\end{figure}

The evolution of structure can be usefully described by a number of
statistical quantities, such as the clumping factor and the velocity
dispersion. These descriptors have been defined in Paper I, and involve the
temporal average of variables such as the density at a given position in the
wind. This is somewhat cumbersome in a moving box model, because a given
position in the wind does not correspond to a unique position within 
the moving box, and calculating a time average requires some awkward 
bookkeeping. 
It is more convenient to replace the temporal averages by spatial averages using
the following ergodic approximation
\begin{equation}
 v_0 \int_{t-T}^t \rho(x,t^{\prime}) {\rm d}t^{\prime} \approx
 \int_{x}^{x+X} \rho(x^{\prime},t) {\rm d}x^{\prime},
\end{equation}
which holds if the statistical properties remain stationary over a time
$\Delta t = L/v_{\rm b}$. We can then calculate the clumping factor and the
velocity dispersion, which are shown in Fig.~\ref{fg:box:stat}. 
The velocity dispersion declines gradually to reach values barely above the
sound speed ($\approx 13$ km/s), although the strongest shock in the final
state still has a jump velocity of 25 km/s (Fig.~\ref{fg:box:final}).
The clumping factor has the oscillating behaviour typical of the competition
between pressure expansion and shell collisions. Between 200 and 1300 $R_{\ast}$,
the clumping factor ranges from 2.5 to 6.

These simulations show that, within the 1D assumption, the wind remains
structured over huge distances. Observational diagnostics such as radio emission 
(both thermal and non-thermal) are inevitably influenced. Inferred mass-loss rates,
that scale as $\sqrt{f_{\rm cl}}$ would be overestimated by a factor of around two.
It appears plausible, again within the approximations, that shocks survive
out to the large distances needed to explain the non-thermal radio emission.

\section{Summary}\label{sect:summary}
We have presented an efficient technique to study the evolution of 
instability-generated structure far away from the star. Because it follows
a small part of the structure, high spatial resolution can be maintained with
a relatively small number of points. Furthermore, the number of depth points
required does not depend on the time-span of the simulation. The small number
of points makes for smaller files and facilitates the analysis of the
results.
As the box
follows the flow, the gas is advected over the grid at much lower speeds than
in a traditional model. This makes {
the condition on the Courant number
much less restrictive,
resulting in a substantial reduction in computation time. The
moving box also has a conceptual advantage, in that it makes it easier to
visualise the physical processes that are happening in the simulation.

We have shown that the high Mach numbers with which the gas moves over the
grid in traditional models
are a cause for concern, as it can cause unphysical structure to appear,
and physical structure to disappear. In this sense, the box model is not only
faster, but also more accurate.  It is therefore a useful tool to check
whether a solution depends on the Galilean frame in which it is obtained.
Using the box model, we have shown that the wind remains clumped out to
$1300 \; R_{\ast}$ and that weak shocks remain present.

A key limitation of the present method is its restriction to just
one-dimension (1D), with focus on the extensive structure in radius, but 
not accounting at all for likelihood that in 2D or 3D models, 
Rayleigh-Taylor or thin-shell instabilities would break up the assumed
azimuthal coherence (Vishniac \cite{Vishniac}).
Future work should thus focus on extending these period box 
techniques to 2D or 3D, in conjunction with recent efforts to develop 
multi-dimensional radiation-hydrodynamics simulations of the initial 
formation of structure arising from the line-deshadowing instability 
(Dessart \& Owocki \cite{Dessart+}; 
Gomez \& Williams \cite{Gomez+}).

\begin{acknowledgements}
We thank John Blondin for the use of VH-1 and for helpful suggestions
regarding the moving shock tube problem. We thank Ronny Blomme for useful
discussions, and Hilde Vanpoucke for preparing some of the figures.
Part of this research was carried out in the framework of the
project IUAP P5/36 financed by the Belgian State, Federal Office for
Scientific, Technical and Cultural Affairs. SPO acknowledges support from NSF
grant AST-0097983.
\end{acknowledgements}

\end{document}